# What is the valence of a correlated solid?  The double life of δ-plutonium

J. H. Shim, K. Haule, and G. Kotliar

*Department of Physics and Astronomy and Center for Condensed Matter Theory, Rutgers University, Piscataway, New Jersey 08854-8019, USA*

While the nuclear properties of the late actinides (Pu, Am, Cm)  are fully understood  and  widely applied to energy generation,  their solid state properties (elastic,  magnetic  and  electronic) do not fit within the standard model of solid state physics and are the subject of active research[1].

Plutonium displays phase transitions with enormous volume differences among its phases and both its Pauli like magnetic susceptibility and resistivity are an order of magnitude larger than those of simple metals[2]. Curium is also highly resistive but its susceptibility is Curie-like at high temperatures and orders antiferromagnetically[3] at low temperatures.  The anomalous properties of the late actinides stem from the competition between the itinerancy and localization of its $f$ electrons, which makes the late actinides elemental strongly correlated materials[4]. A central problem in this field is to understand the mechanism by which these materials resolve these conflicting tendencies.

In this letter we identify the electronic mechanisms responsible for the anomalous behaviour of late actinides. We revisit the concept of valence using theoretical approach that treats magnetism, Kondo screening, atomic multiplet effects, spin orbit coupling and crystal field splitting on the same footing. Plutonium is found to be in a rare mixed valent state, namely its ground state is a **superposition of two distinct valencies. Curium settles in a single valence magnetically ordered state at low temperatures. The $f^7$ atomic configuration of Curium is contrasted with the multiple configuration manifolds present in Plutonium ground state which we characterize by a valence histogram. The balance between the Kondo screening and magnetism is determined by the competition between spin orbit coupling and the strength of atomic multiplets which is in turn regulated by the degree of itinerancy.**

**The approach presented here, highlights the electronic origin of the bonding anomalies in plutonium and can be applied to predict generalized valences and the presence or absence of magnetism in other compounds starting from first principles.**

To understand unique properties of the elemental plutonium and its distinctive placement in the periodic table, it is cardinal to build a theory of actinides that can describe itinerant actinides as well as late actinides beyond plutonium. Curium, which follows americium in periodic table, provides a very useful analog to plutonium expanded beyond the equilibrium volume of its δ phase. Curium is obtained by adding an electron to americium inert *5f* shell (J=0), while plutonium is obtained by creating a hole in the shell. The ability to predict magnetism in curium and its absence in plutonium is a strong test of the methodology, and we will use it as a benchmark for the quality of our theory of actinides.

To treat realistic mutliplets and band structure we use a new implementation of the merger of Local Density approximation and Dynamical Mean Field method (LDA+DMFT)[5] with an accurate impurity solver, vertex corrected one-crossing approximation[5] which we have further crosschecked against a continuous time Quantum Monte Carlo method[6]. We treat the full atomic physics (all the Slater integrals $F_0$, $F_2$, $F_4$, and $F_6$) on the same footing with the realistic band structure obtained by the relativistic version of the linear muffin-tin orbital method[7]. This LDA+DMFT approach treats magnetism and Kondo physics on an equal footing and takes into account all the multiplet structure and crystal fields. As in all the earlier studies[8] we take Coulomb interaction $U = 4.5$ eV and we compute the rest of the Slater integrals ($F_2$, $F_4$, and $F_6$) from atomic physics[9].

Underlying the Dynamical Mean Field approach[5] is a physical picture in which the f electrons are fluctuating among the different atomic configurations and exchanging electrons with a reservoir. The properties of reservoir are determined self consistently from the knowledge of the local spectral function. The self consistency condition allows solutions with partially delocalized f electrons forming quasiparticle bands, but at the same time, the f electrons are allowed to preserve their atomic character at short time, which leads to formation of Hubbard bands in the spectral function. When the f electrons get sufficiently localized, magnetic solutions are possible and become more energetically favorable than paramagnetic solutions at low temperature.

The fingerprint of strong correlations is encoded in the many body self-energy, which becomes a $14 \times 14$ matrix dependent on frequency but not momentum. Added to the Hamiltonian matrix, which also contains *spd* electrons, gives rise to Green's function of the problem, $G(\omega, \mathbf{k})$. Integrated over momentum, it results in the local spectral function, which is measured in photoemission and inverse photoemission experiments.

To investigate phases with antiferromagnetic long range order, self energy is allowed to be spin dependent, and the lattice is partitioned into two different sublattices,

*A* and *B*. Because different sites experience different environments, they are affected by different hybridization functions. We start from a magnetic state where the effective mediums for the different projections of the *z* component of the electron total angular momentum $j_z$ at the *A* and *B* sublattices are different, and we watch how they evolve under iteration. We will see that plutonium does not take advantage of this possibility and hence is non-magnetic, while curium does.

Starting from a general initial condition, upon iteration the plutonium spectral functions converge to a $j_z$ independent spectral function. On the other hand, curium has 14 nonequivalent spectral functions, reflecting the low temperature antiferromagnetism, as displayed in the figure 1. Notice the Hubbard bands, atomic-like features, and the Kondo resonance, which is present in plutonium but not in curium. Multiplet effects are clearly visible and provide widths to the Hubbard bands, as pointed out previously in studies of americium[13]. Here we show that they play an even more important role in determining the renormalized Fermi energy, or the Kondo scale. For plutonium we obtain a Kondo energy of the order of 800 K, which compares favorably with the measured specific heat of the order of 60 mJ/molK[14]. Turning off the Hund's rule coupling, which is possible theoretically but not experimentally, would result in a much larger Kondo energy in plutonium and in a non-magnetic heavy fermion state in Cm in disagreement with experiments. Hence the Hunds rule coupling plays an unexpected role in the actinide series, renormalizing down the Kondo energy.

To define valence we focus on the reduced density matrix of the *f* states at a given site, which is obtained from the exact density matrix of the solid by tracing over all degrees of freedom except for those of the 5*f* shell at a given site. The eigenvalues of this reduced density matrix give the probability of observing different *f* electron atomic configurations at a given unit cell associated with an actinide nucleus. The solution of the DMFT impurity model allows us to visualize the *f* electrons as fluctuating between various atomic configurations and exchanging electrons with the surrounding medium. As a function of time, the *f* electrons in the atom change their atomic configuration while absorbing and emitting electrons into the bath. We keep track of the different atomic configurations visited and draw them as histograms, which give complementary information to the photoemission spectra. These histograms for δ-plutonium and curium are presented in figure 2. Notice that plutonium does not have a well defined valence, its *f* electrons live a double life, spending considerable time in several atomic configurations, even though its average *f* electron count is close to $5f^5$. We describe this situation with a histogram which is peaked for a few atomic eigenstates, including atomic ground states of $5f^5$ and $5f^6$. The system is in a mixed valence state[15] with an average *f* occupation of $n_f \sim 5.2$. In curium the *f* electrons are locked into one $5f^7$ dominant configuration, and the histogram is peaked only for the ground state of the atom.

X-ray absorption from the core *4d* states is a powerful probe of the valence. The strong spin orbit coupling of the core states gives rise to two spin-orbit split absorption lines, representing $4d_{5/2} \rightarrow 5f$ and $4d_{3/2} \rightarrow 5f$ transitions [16,17]. The branching ratio *B*, i.e. the relative strength of the $4d_{5/2}$ absorption is a measure of the strength of the spin-orbit coupling interaction in the *f* shell. Ignoring the electrostatic interaction between core and valence electrons, known to be negligible in plutonium[18], results in the following general expression obtained first by Van der Laan and collaborators[16]:

$$B = \frac{A_{5/2}}{A_{5/2} + A_{3/2}} = \frac{3}{5} - \frac{4}{15} \frac{1}{14 - <n_{5/2}> - <n_{7/2}>} \sum_{i \in f} < \vec{l}_i \cdot \vec{s}_i > \quad (1)$$

$$\sum_{i \in f} < \vec{l}_i \cdot \vec{s}_i > = \frac{3}{2} <n_{7/2}> - 2 <n_{5/2}>$$

where $A_{5/2}$ and $A_{3/2}$ are associated with area under the peaks corresponding to $4d_{5/2} \rightarrow 5f$ and $4d_{3/2} \rightarrow 5f$ transitions, respectively. The term $\sum_{i \in f} < \vec{l}_i \cdot \vec{s}_i >$ measures the strength of the spin-orbit coupling in the valence $f$ states, and $n_{5/2}$ and $n_{7/2}$ are averaged partial occupations of the $f$ valence states.

Comparison of measured branching ratios in actinides[19] with atomic physics computations[18] indicated the important role of spin orbit coupling and gave a strong evidence in favor of a $5f^5$ configuration. We now evaluate B for both curium and plutonium, using DMFT, which goes beyond atomic physics by incorporating the effects of itinerancy and multiple valences. The DMFT results are summarized in table 1 and are compared with experiment[19] where available. We stress the importance of first principles calculations including spin-orbit coupling, itinerancy and multiplet effects, since these are competing effects. While the spin orbit coupling increases faster than the Hund's coupling (Slater $F_2$) with increasing number of $f$ electrons, the changes in the degree of itinerancy are of comparable magnitude and affect the relative strength of the effective spin-orbit coupling to the Hund's coupling and crystal fields. As a result the effective spin-orbit coupling is smaller in curium than plutonium, placing curium much closer to the Russell-Saunders (LS) coupling than plutonium[12,20] (see last column in table 1). The resulting curium moment $\mu \sim 2\sqrt{S(S+1)} \sim 7.9$ $\mu_B$ is close to the observed experimental value[3], which is clearly incompatible with $j$-$j$ coupling (one $f_{7/2}$ electron would result in a moment of $l+2s = 3+1$ $\mu_B$).

We now contrast our findings with earlier approaches. Spin density functional theory calculations (SDFT) consistently predict that all late actinides (Pu, Am, and Cm) are magnetic with a large ordered magnetic moments of the order of a few Bohr magnetons[21]. Experimentally, however, it is now established that no fluctuating or ordered moments exist in metallic plutonium[2] and americium[22] but large moment is seen in Curium. SDFT can be viewed as a form of static mean field theory, which is known to produce spurious magnetic states in order to mimic correlations.

Other theories of plutonium posits that in this material some *5f* electrons are localized and some are itinerant. The mixed level model[23] assumes that 4 electrons are localized, i.e., condensed into an atomic like singlet, while one *f* electron is itinerant.  In the self-interaction corrected LDA, the valence of the *f* is frozen, and the total energy is determined for each valence to select the one with the lowest total energy. The later approach[24] finds that configurations with four three, two, one or zero localized *5f* electrons are almost degenerate. This can be taken as an indication, that the dynamical treatment of valence presented in this paper is needed for Plutonium.

Other calculations (LDA+U[25], DMFT-FLEX[26]) suggest that plutonium is close to inert $5f^6$ configuration with singlet formed out of 6 localized *f* electrons. A non magnetic configuration naturally accounts for the absence of moments in plutonium but is too inert to account for the fact that the specific heats in $\alpha$ and $\delta$ phases differ by more than a factor of two[2,27]. Furthermore, X-ray absorption experiments[28] and photoemission on thin plutonium layers[29] as well as previous DMFT calculations for plutonium[8] are very suggestive that the *f* electrons are close to $5f^\delta$ configuration.

The full dynamic treatment of multiplets and Kondo physics, carried out in this paper, brings a significant admixture of $5f^6$ valence establishing continuity with weak coupling treatments[26], while accounting  for the mass enhancement in alpha and delta Pu. Our new technical and conceptual advances in understanding $\delta$-Pu and Cm lead to several experimental predictions. In the paramagnetic state, the volume enclosed by the Fermi surface of Pu should contain and even number of electrons, while that of Cm should contain an odd number of electrons, i.e., his 3 *spd* electrons.  These predictions should be tested with de Haas-van Alphen experiments or angle resolved photoemission experiments. Furthermore, the physical picture of plutonium as a mixed valence metal provides a natural explanation for the large sensitivity of its volume to small changes in temperature, pressure or doping. Moreover, the mixed valence nature of Pu can be probed by inverse photoemission experiments and by optical conductivity experiments, which should display a hybridization gap on a scale corresponding to several times the Kondo energy. In addition to the standard low energy Drude peak, the optical conductivity should display a hybridization dip around 1000 cm$^{-1}$ and a broad mid infrared peak between 3000 and 4000 cm$^{-1}$. The one electron spectra and the X-ray branching ratio of curium are further quantitative predictions, which can be tested experimentally via photoemission and X-ray absorption measurements.

Finally, the use of DMFT for extracting valence histograms and thinking about mixed valences should have applications for many other strongly correlated compounds. Two pressing examples are $UO_2$ and $PuO_2$, important byproducts in nuclear reactors whose valence is not well understood.

**Acknowledgements**: This work was supported by the DOE and Korean Research Foundation Grant

funded by the Korean Government (MOEHRD) (KRF-2005-214-C00191).

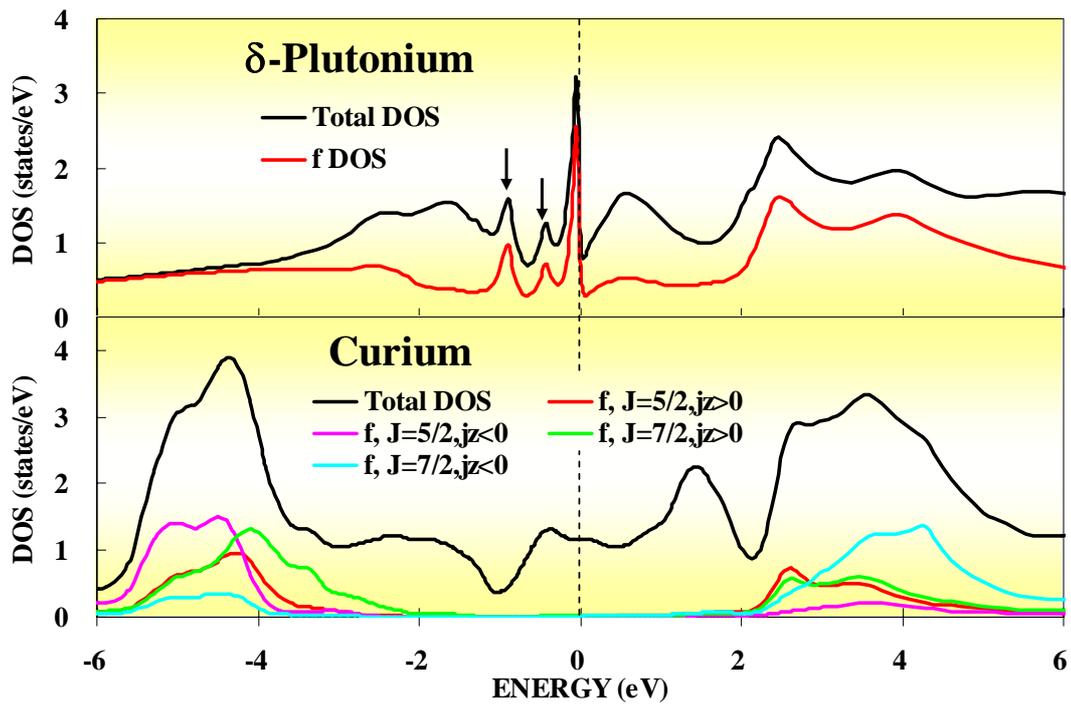

**Figure 1: The spectral functions of δ-plutonium and FCC curium.** The δ-plutonium is in paramagnetic state with its moment totally screened by the Kondo effect, which is observed as a resonance at the Fermi level in the *f* spectra. In addition to the broad Hubbard band and sharp Kondo peak, two additional peaks below the Fermi level appear in our spectral function (marked with arrows) which were recently identified in photoemission experiments[10,11,12]. The virtual charge fluctuations that give rise to Kondo peak are primarily between the blue central peak in figure 2 (ground state of $N_f$=5) to the green side peak |$N_f$=6,*J*=0,γ=0> (ground state of $N_f$=6) in the same panel. The two subbands around 0.5 and 0.85 eV, indicated by arrows, come mostly from the charge fluctuations between the second blue peak |$N_f$=5,*J*=7/2,γ=0> to the ground state of $N_f$=6 and |$N_f$=5,*J*=5/2,γ=1> to the same ground state of $N_f$=6, respectively. They disappear above the coherence temperature therefore they are part of the coherent many body spectra. The second panel shows the curium spectral function in the antiferromagnetic state. The diagonal

components of the 14×14 matrix spectral function are shown separately for the positive and negative components of the electron spin $j_z$.

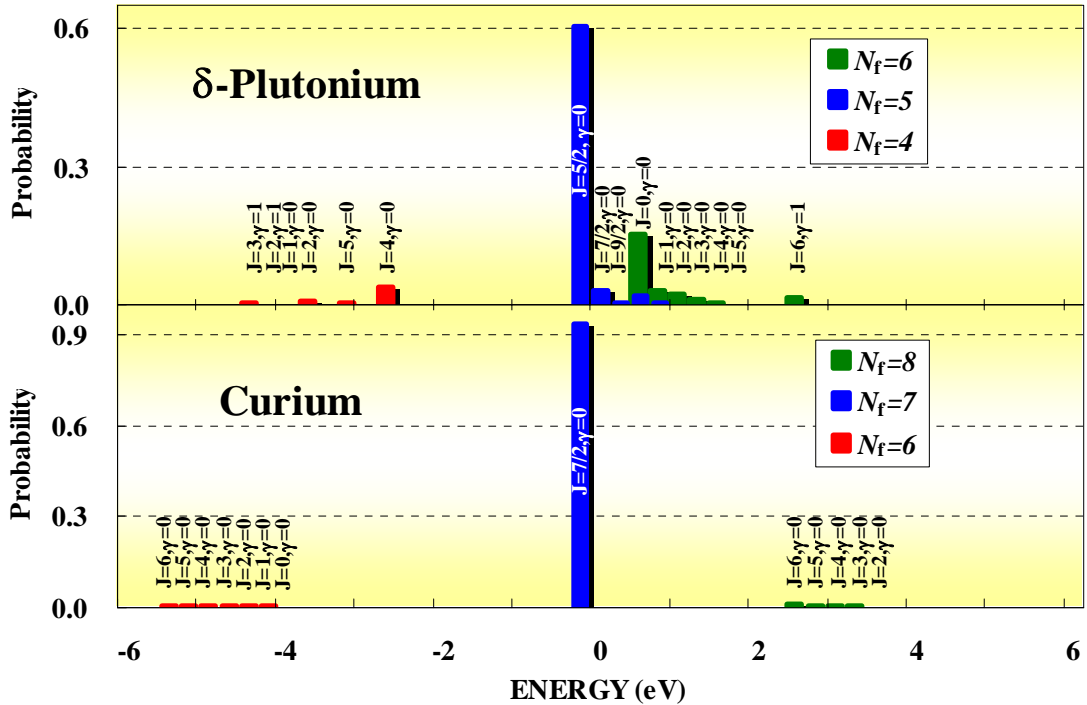

**Figure 2: Projection of the DMFT ground state to various atomic configurations.** The histograms describe the generalized concept of valence, where the $f$ electron in the solid spends appreciable time in a few atomic configurations. The height of the peak corresponds to the fraction of the time the $f$ electron of the solid spends in one of the eigenstates of the atom, denoted by the total spin $J$ of the atom. We summed up the probabilities for the atomic states which differ only in the $z$ component of the total spin $J_z$. The rest of the atomic quantum numbers are grouped into a single quantum number $\gamma = 0, 1, \ldots$. The $x$ axis indicates the energy of atomic eigenstates in the following way: Energy($N_f$-1,$J\gamma$)=$E_{atom}$($N_f$,ground-state)-$E_{atom}$($N_f$-1,$J\gamma$) and Energy($N_f$+1,$J\gamma$)= $E_{atom}$($N_f$+1,$J\gamma$)-$E_{atom}$($N_f$,ground-state), where $N_f$ is 5 and 7 for $\delta$-plutonium and curium, respectively.

**Table 1 The *f*-electron count and branching ratio, *B*, of the $N_{4,5}$ edge spectra of δ-plutonium and curium. $B_{LS}$ and $B_{jj}$ correspond to limiting cases of the pure Russell-Saunders and *j-j* coupling, respectively.**

|  | *f* count | $B_{DMFT\ theory}$ | $B_{exp}$[19] | $B_{LS}$[18] | $B_{jj}$[18] | $(B_{DMFT} - B_{LS})/(B_{jj} - B_{LS})$ |
|---|---|---|---|---|---|---|
| δ-Plutonium | 5.2 | 0.83 | 0.847 | 0.69 | 0.90 | 0.67 |
| Curium | 7.0 | 0.75 |  | 0.6 | 1.0 | 0.38 |